\documentclass[american,aps,pra,reprint, superscriptaddress]{revtex4-1}
\usepackage{color}
\usepackage{times}
\usepackage{subfig}
\usepackage{bm}
\usepackage{graphicx}
\usepackage{graphics}
\DeclareGraphicsExtensions{.pdf,.png,.jpg}
\usepackage{amsbsy}
\usepackage{amsmath}
\usepackage{amsfonts}
\usepackage{amsthm}
\usepackage{float}
\usepackage{amssymb}
\usepackage{textcomp}
\usepackage{mathtools}
\usepackage[subnum]{cases}

\definecolor{purple(html/css)}{rgb}{0.5, 0.0, 0.5}
\newcommand{\ket}[1]{| #1 \rangle}

\newcommand{\ketbra}[2]{| #1 \rangle \langle #2 |}

\newcommand{\mi}{\mathrm{i}}
\begin{document}

\title{Binegativity of two qubits under noise}
\author{Sk Sazim}
 \email{sk.sazimsq49@gmail.com}
 \affiliation{QIC group, Harish-Chandra Research Institute, HBNI, Allahabad, 211019, India}
\author{Natasha Awasthi}
\affiliation{Govind Ballabh Pant University of Agriculture and Technology, Pantnagar, Uttarakhand 263145, India}
\begin{abstract}
Recently, it was argued that the binegativity might be a good quantifier of 
entanglement for two-qubit states. Like the concurrence and the negativity, 
the binegativity is also analytically computable quantifier for all two qubits. Based on numerical evidence, 
it was conjectured that it is a PPT (positive partial transposition) monotone and  
thus fulfills the criterion to be a good measure of entanglement. 

In this work, we investigate its behavior under noisy 
channels which indicate that the binegativity is decreasing monotonically with respect to increasing noise. We also find that the binegativity is closely connected to the negativity and has closed analytical form for arbitrary two qubits. 
Our study supports the conjecture that the binegativity is a monotone. 
\end{abstract}

\maketitle
\section{Introduction}
Quantum entanglement is a fundamental non-classical feature of multiparticle quantum systems. It is a key resource for 
many quantum information processing tasks. Hence, characterizing (witnessing as well as quantification) of entanglement 
is of immense importance. 

In the last two decades, substantial amount of progress has been made in characterizing entanglement of two-qubit 
systems \cite{entangr}. Although the entanglement structure of pure bipartite systems is well understood, 
much attention is required 
to fully understand it for mixed two-qubit states \cite{entangr}. Quantification of entangled state is related with the 
inconvertibility 
between entangled states under local operations and classical communications (LOCC), i.e., the quantities which do not 
increase under LOCC are the entanglement quantifiers \cite{Locc3,Locc1,Locc2,Locc4}. Finding such measures are 
important for better understanding of the
entangled states \cite{Locc3,EntMea3,EntMea2,EntMea1}. Out of many extant entanglement quantifiers, the concurrence 
\cite{Conc1,Conc2} and the negativity \cite{Negat1} are easily computable for 
two-qubit mixed states. Although, negativity and concurrence coincide for pure two qubit states, they produce different 
ordering for mixed states \cite{Order1}.

One breakthrough discovery in entanglement theory is Peres-Horodecki criteria \cite{PPT1,PPT2}. 
They found that using partial transposition 
operations one can detect entanglement in composite quantum systems. Let us consider a bipartite system 
$\rho$, then its partial transposition in one of the subsystems is defined as $\rho^{\Gamma}$. The state satisfying 
$\rho^{\Gamma}\geq 0$ are called positive under partial transposition (PPT states). It is well known that all the PPT 
states of two qubits are separable states. The negativity captures the degree of violation of PPTness in the 
two-qubit states and it is an entanglement monotone \cite{EntMea1}. Note that there exist no 
known physical interpretation of partial 
transposition operations. 
The negativity can be expressed as 
\begin{equation}
 N(\rho)=2{\rm Tr}[\rho^{\Gamma_-}]=\parallel \rho^{\Gamma} \parallel_1-1,
\end{equation}
where $\parallel \cdot \parallel_1$ denotes trace-norm and we follow the notation 
$\rho^{\Gamma_-}=(\rho^{\Gamma})_-$ to denote the negative component of $\rho^{\Gamma}$. (It is defined in Eq.(\ref{neg_comp}).)

In Ref.\cite{bineg}, authors discussed a computable quantity called `the binegativity' which may be considered as 
a potential entanglement 
measure. The concept of binegativity was first introduced in the context of 
relative entropy of entanglement \cite{binegWhy1}. 
It was shown that if $|\rho^{\Gamma_-}|^{\Gamma_-}\geq 0$, the asymptotic relative entropy of entanglement with
respect to PPT states does not exceed the so-called Rains bound \cite{PPTneg,binegWhy1}, where $|\rho|=\sqrt{\rho.\rho}$. 
This condition also guarantees 
that the PPT-entanglement cost
for the exact preparation is given by the logarithmic negativity \cite{binegWhy2, binegWhyAl} which provides the operational 
meaning to logarithmic negativity \cite{LogNeg}.
The binegativity for two-qubit state is given by \cite{bineg}
\begin{eqnarray}
 N_2(\rho)&=&{\rm Tr}[\rho^{\Gamma_-}]+2{\rm Tr}[\rho^{\Gamma_-\Gamma_-}],\nonumber\\
 &=& \frac{1}{2} N(\rho)+2 {\rm Tr}[\rho^{\Gamma_-\Gamma_-}],
\end{eqnarray}
where $\rho^{\Gamma_-\Gamma_-}=((\rho^{\Gamma_-})^{\Gamma})_-$. The binegativity has similar properties like 
negativity in two qubit systems while the former may not be a monotone under both LOCC and PPT channels
\cite{LogNeg,PPTneg,bineg,note12}. On the basis of 
numerical evidence, it is conjectured that the binegativity behaves 
monotonically under both LOCC and PPT channels \cite{bineg}. 
Based on 
this conjecture, the binegativity might be identified as a valid measure of entanglement for 
two qubit states. The binegativity has following properties \cite{bineg}:
\begin{enumerate}
 \item It is positive always and vanishes for two-qubit separable states.
 \item It is invariant under local unitary operations.
 \item For all two qubit states $N_2(\rho)\leq N(\rho)\leq C(\rho)$ and $N_2(\rho)=N(\rho)$ if $N(\rho)=C(\rho)$. In 
particular for all pure two qubit states, $\ket{\psi}$, $N_2(\ket{\psi})= N(\ket{\psi})= C(\ket{\psi})$, 
where $C$ denotes concurrence.
\end{enumerate}

The comparison between the negativity and the concurrence have been studied extensively and these measures 
give different order for the two qubit states, as there exists different states with equal concurrence but different 
negativity and vice versa \cite{EntMea2,Order1,order2,order3}. The binegativity also gives a unique orderings 
of two-qubit states \cite{bineg}.  
There exists some two qubit states with same negativity and same concurrence 
but have different values of binegativity. 
All these findings indicate that
the binegativity may be a new member in the set of extant entanglement quantifier.

In this work, 
we study its behavior 
under noisy channels, 
specifically, under amplitude damping (AD), phase damping (PD) and depolarizing (DP) channels and find that 
it is decreasing monotonically with the increasing noise. We also observe that the behavior of 
the binegativity is quite similar under noisy channels. All these study indicates that the bonafied measure, 
the binegativity, might be a entanglement monotone.

In the next section, we establish a functional relation between the binegativity and the negativity. 
We also discuss the behavior of the binegativity under twirling operation. Then we calculate the 
binegativity for some class of states in section-III. In section-IV, we study the behavior of the binegativity under 
the noisy channels. We conclude in the last section.
\section{Binegativity -- a LOCC monotone?}
Although we do not have a proof for monotonicity of the binegativity under LOCC/PPT, we will 
address the issue to some extend. Mainly we will show that the binegativity contains a nontrivial term 
which may increase under some local operations but on average the binegativity is not increasing. Here we 
focus our numerical study only for twirling operations.

Binegativity of two qubit state $\rho$ can explicitly be expressed in terms of negativity. \\
\noindent \textbf{Lemma.}-- \textit{The binegativity, $N_2(\rho)=\frac{1}{2}N(\rho)\big[1+N(\rho_{\psi})\big]$, where 
$\rho_{\psi}=\ketbra{\psi}{\psi}$ with $\ket{\psi}$ being the normalized eigen vector corresponding to 
the negative eigen value of $\rho^{\Gamma}$.}\\
\noindent \textit{Proof}.- It is well known that the partial transposition of any two qubit entangled state 
has exactly one negative eigenvalue, and the eigenstate (pure) corresponding to it must be an entangled state. 
Hence the negative component of $\rho^{\Gamma}$ is of the form 
\begin{equation}
 \rho^{\Gamma_-}={\rm Tr}\big[\rho^{\Gamma_-}\big]\rho_{\psi},
 \label{neg_comp}
\end{equation}
where $\rho_{\psi}=\ketbra{\psi}{\psi}$ with $\ket{\psi}$ being the normalized eigen vector corresponding to 
the negative eigen value of $\rho^{\Gamma}$. Now the form of $\rho^{\Gamma_-\Gamma_-}$ is given by
\begin{equation}
 \rho^{\Gamma_-\Gamma_-}={\rm Tr}\big[\rho^{\Gamma_-}\big]\rho_{\psi}^{\Gamma_-}.
\end{equation}
Hence, ${\rm Tr}\big[\rho^{\Gamma_-\Gamma_-}]=\frac{1}{4}N(\rho)N(\rho_{\psi})$. Therefore the binegativity can be 
expressed as follows 
\begin{eqnarray}
 N_2(\rho)=\frac{1}{2}N(\rho)\big[1+N(\rho_{\psi})\big].
\end{eqnarray}
Hence the proof. $\square$

With the above expression, we can conclude that the binegativity and the negativity are related quantities. The binegativity 
and negativity coincide for two qubit pure states as in this case $\rho_{\psi}$ is a maximally entangled state. In fact, 
it is true for Werner states also. 


%

We know that the negativity is a monotone 
under PPT operations \cite{EntMea1, LogNeg,PPTneg}. Having close resemblance with negativity, one 
might also expect that the binegativity is a monotone. 
However in Ref.\cite{bineg}, based on numerical evidence, it was conjectured that the binegativity might be a PPT monotone. 
Analytically, it is hard to prove the monotonicity of the binegativity 
because of the presence of the term like $N(\rho_{\psi})$. For example, 
any two qubit entangled state can be transformed to a less entangled Werner state by twirling operations \cite{twirl_wer} and 
for the Werner state, $\rho_{\psi}$ is maximally entangled i.e., $N(\rho_{\psi})=1$. Therefore, although the 
overall entanglement is decreasing the contribution from the term, $N(\rho_{\psi})$ may increase.

In \cite{twirl_wer}, Werner showed that any state $\rho$ can be transformed to a Werner state by applying the twirling operator:
\begin{equation}
 \rho_{Wer}=\int dU(U\otimes U)\rho (U\otimes U)^{\dagger},
\end{equation}
where integral is performed with respect to Haar measure on the unitary group, $U(d)$. This operation can transform 
any entangled state to a less entangled Werner state. Therefore, under the twirling the binegativity should also decrease for 
two qubit case. \textit{We have 
numerically checked that the binegativity is indeed monotonically decreasing under twirling.}

Now we will compute the binegativity for some class of states.
\section{Binegativity of some class of states}
Here we will compute the binegativity for some two qubit mixed states. For example, 
we will consider the following states:

\noindent \textbf{Werner state}: The Werner state is $U\otimes U$ invariant state. A two qubit 
Werner state is given by
\begin{equation}
 \rho_{Wer}=\frac{1-p}{4}\mathbb{I}_4+p\ketbra{\psi^-}{\psi^-},
\end{equation}
where $\ket{\psi^-}=\frac{1}{\sqrt{2}}(\ket{01}-\ket{10})$ is the singlet state and $ p\in [0,1]$ is the 
classical mixing. The state is entangled for $p>\frac{1}{3}$. For this state the concurrence, the negativity and the 
binegativity are same and are equal to $\frac{3p-1}{2}$ for $p>\frac{1}{3}$.

\noindent \textbf{Bell diagonal states}: The Bell diagonal states can be expressed in canonical form as
\begin{equation}
 \rho_{Bell}=\frac{1}{4}(\mathbb{I}_4+\sum_ic_i\sigma_i\otimes\sigma_i),
\end{equation}
where $c_i\in [-1,1]$. The state, $\rho_{Bell}$ is a valid density matrix if its eigen values $\lambda_{mn}\geq 0$, 
where $\lambda_{mn}=\frac{1}{4}[1+(-1)^mc_1-(-1)^{m+n}c_2+(-1)^nc_3]$ with $m,n=0,1$. For this state, 
the concurrence, the negativity and 
the binegativity are equal to $2\lambda_{\max}-1$, where $\lambda_{\max}$ is the maximum eigenvalue of $\rho_{Bell}$.

\noindent \textbf{MEMs}: The two qubit maximally entangled mixed states (MEMs) are the most 
entangled states for a 
given mixedness \cite{Mems}. These states with concurrence $C$ are 
\begin{equation}
 \rho_{MEM}=\begin{pmatrix}
g(C) & 0 & 0 & \frac{C}{2} \\
0 & 1-2g(C) & 0 & 0 \\
0 & 0 & 0 & 0 \\
\frac{C}{2} & 0 & 0 & g(C) \\
\end{pmatrix},
\end{equation}
where $g(C)$ is equal to $\frac{C}{2}$ for $C\geq \frac{2}{3}$ and $\frac{1}{3}$ for $C< \frac{2}{3}$. The negativity of this 
state is given by 
\begin{equation}
 N(\rho_{MEM})=\left\{
     \begin{array}{@{}l@{\thinspace}l}
       \sqrt{(1-C)^2+C^2}-(1-C)  &: \text{if}\: C\geq \frac{2}{3},\\
       \frac{1}{3}(\sqrt{1+9C^2}-1) &: \text{if}\: C< \frac{2}{3}. \\
     \end{array}
   \right.
\end{equation}
The negativity of $\rho_{MEM}$ for $C\geq \frac{2}{3}$ will never exceed its concurrence $C$ 
\cite{order3}. The binegativity of this 
state can be simplified to 

\begin{equation}
 N_2(\rho_{MEM})=\left\{
     \begin{array}{@{}l@{\thinspace}l}
       \frac{N(\rho_{MEM})}{2}\left[1+\frac{C}{\sqrt{(1-C)^2+C^2}}\right]  &: \text{if}\: C\geq \frac{2}{3},\\
       \frac{N(\rho_{MEM})}{2}\left[1+\frac{3C}{\sqrt{1+9C^2}}\right] &: \text{if}\: C< \frac{2}{3}. \\
     \end{array}
   \right.
\end{equation}
Later a more general MEMs were considered in Ref. \cite{gMems} which are expressed as 
\begin{equation}
 \rho_{gMEM}=\begin{pmatrix}
x+\frac{\gamma}{2} & 0 & 0 & \frac{\gamma}{2} \\
0 & a & 0 & 0 \\
0 & 0 & b & 0 \\
\frac{\gamma}{2} & 0 & 0 & y+ \frac{\gamma}{2}\\
\end{pmatrix},
\label{gMemstates}
\end{equation}
where $x,y,a,b,\gamma\geq 0$. The subset of these states (i.e., $x=0=y=b$, $\gamma=C$ and $a=1-C$) are MEMs. 
The concurrence and the negativity of these states (Eq.(\ref{gMemstates})) are given by 
$C(\rho_{gMEM})=\max[0,\gamma-2\sqrt{ab}]$ and $N(\rho_{gMEM})=\sqrt{(a-b)^2+\gamma^2}-(a+b)$ respectively. These results 
show that the state $\rho_{gMEM}$ is entangled when $\gamma>2\sqrt{ab}$ or $\sqrt{(a-b)^2+\gamma^2}>(a+b)$. The binegativity of 
this state simplifies to 
\begin{equation}
 N_2(\rho_{gMEM})=\frac{N(\rho_{gMEM})}{2}\left[1+\frac{\gamma}{\sqrt{(a-b)^2+\gamma^2}}\right].
\end{equation}
\section{Binegativity under noisy channels}
Entanglement is inevitably fragile when exposed to noise. The phenomenon is called decoherence 
\cite{channel3,channel4,channel5}. There exists 
several models to describe the different types of noises (effect of environment on systems). These models are known 
as quantum channels \cite{channel1,channel2}. Mathematically, channels are completely positive trace preserving (CPTP) maps 
having operator sum representations \cite{channel5}. Three important classes of channels are amplitude damping (AD) channels, 
phase damping (PD) channels and depolarizing (DP) channels \cite{channelBoo}. First we will briefly discuss these channels.\\
\noindent \textbf{AD}: The Kraus operator (operator-sum) representation of AD channels are 
$$K_0=\begin{pmatrix}
1 & 0 \\
0 & \sqrt{1-\eta}
\end{pmatrix}\quad\quad\mbox{and}\quad\quad K_1=\begin{pmatrix}
0 & \sqrt{\eta} \\
0 & 0
\end{pmatrix},$$
where $K_0^{\dagger}K_0+K_1^{\dagger}K_1=1$. The evolution of a qubit density matrix $\varrho$ under this channel is 
given by 
$\varrho\mapsto \varrho\prime=K_0 \varrho K_0^{\dagger}+K_1 \varrho K_1^{\dagger}$.
\\
\noindent \textbf{PD}: The PD channels plays an important role in the transition from quantum to classical world 
\cite{channelBoo}. The 
Kraus operators to represent a PD channels are --
$K_0=\sqrt{1-\eta}\mathbb{I}_2$, 
$$K_1=\begin{pmatrix}
\sqrt{\eta} & 0 \\
0 & 0
\end{pmatrix}\quad\quad\mbox{and}\quad\quad K_2=\begin{pmatrix}
0 & 0 \\
0 & \sqrt{\eta}
\end{pmatrix}.$$
Under this channel a qubit state is transformed to 
$ \varrho\mapsto \varrho\prime=(1-\eta)\varrho+ \eta\varrho_{dia}$,
where $\varrho_{dia}$ is the density matrix with diagonal elements of $\varrho$. Hence, under this channel off-diagonal 
elements of the density matrix decreases with time.\\
\noindent \textbf{DP}: The operator-sum representation of the DP channel is 
$$K_0=\sqrt{1-\eta}\mathbb{I}_2\quad\quad\mbox{and}\quad\quad K_i=\sqrt{\frac{\eta}{3}}\sigma_i,$$ 
where $i=1,2,3$. A qubit density matrix under DP channel transform to
$ \varrho\mapsto \varrho\prime=(1-\eta)\varrho+ \frac{\eta}{3}\sum_i \sigma_i\varrho\sigma_i$,
where $0 \leq \eta \leq 1$.

We study the effect of the above mentioned channels for a mixed state given below
\begin{equation}
 \rho_{EW}=\frac{1-p}{4}\mathbb{I}_4+p\ketbra{\Psi}{\Psi},
\end{equation}
where $\ket{\Psi}=\alpha\ket{00}+\beta\ket{11}$ with $|\alpha|^2+|\beta|^2=1$. To observe the environmental effects, 
we will consider two type of channels -- $\Lambda_{i0}=K_i\otimes \mathbb{I}_2$ and $\Lambda_{ij}=K_i\otimes K_j$.  
The state generated due to the application of channels on one particle ($\Lambda_{i0}$) 
and both particles ($\Lambda_{ij}$) of the target state are discussed in details.

\begin{figure}[h]
\centering
\includegraphics[scale=0.7]{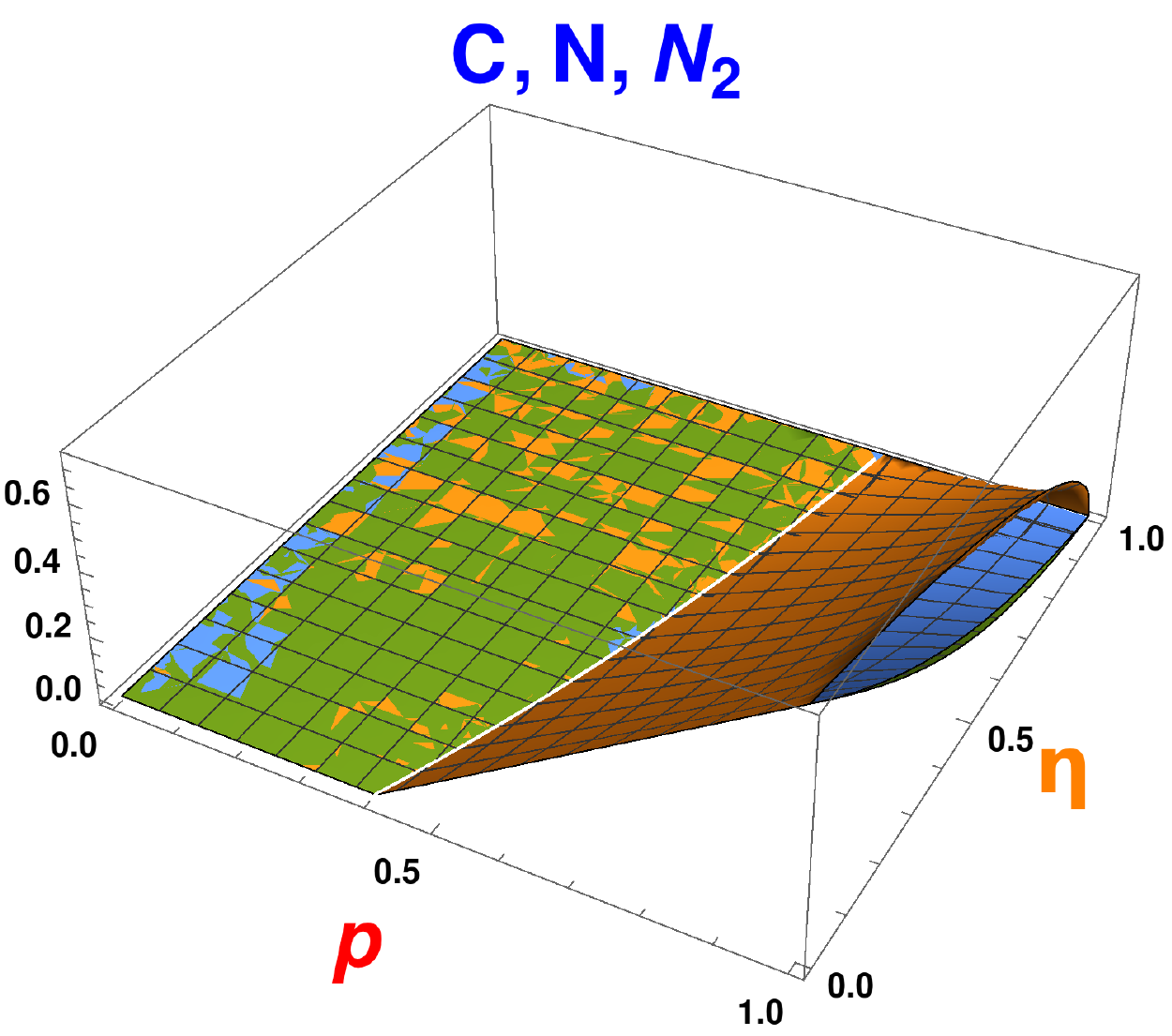}
\caption{(Color online) The figure shows the behavior of the concurrence ($C$), 
the negativity ($N$), and the binegativity ($N_2$) of 
initial state $\rho_{EW}$ versus the mixing parameter $p$ and the channel parameter $\eta$ for $\alpha=0.4$ under 
the action of AD channel on one particle of the state. 
It depicts that the concurrence (orange) is more robust than the rest while the binegativity (green) is the most fragile 
under AD channel.}\label{AD1}
\end{figure}

\begin{figure}[h]
\centering
\includegraphics[scale=0.6]{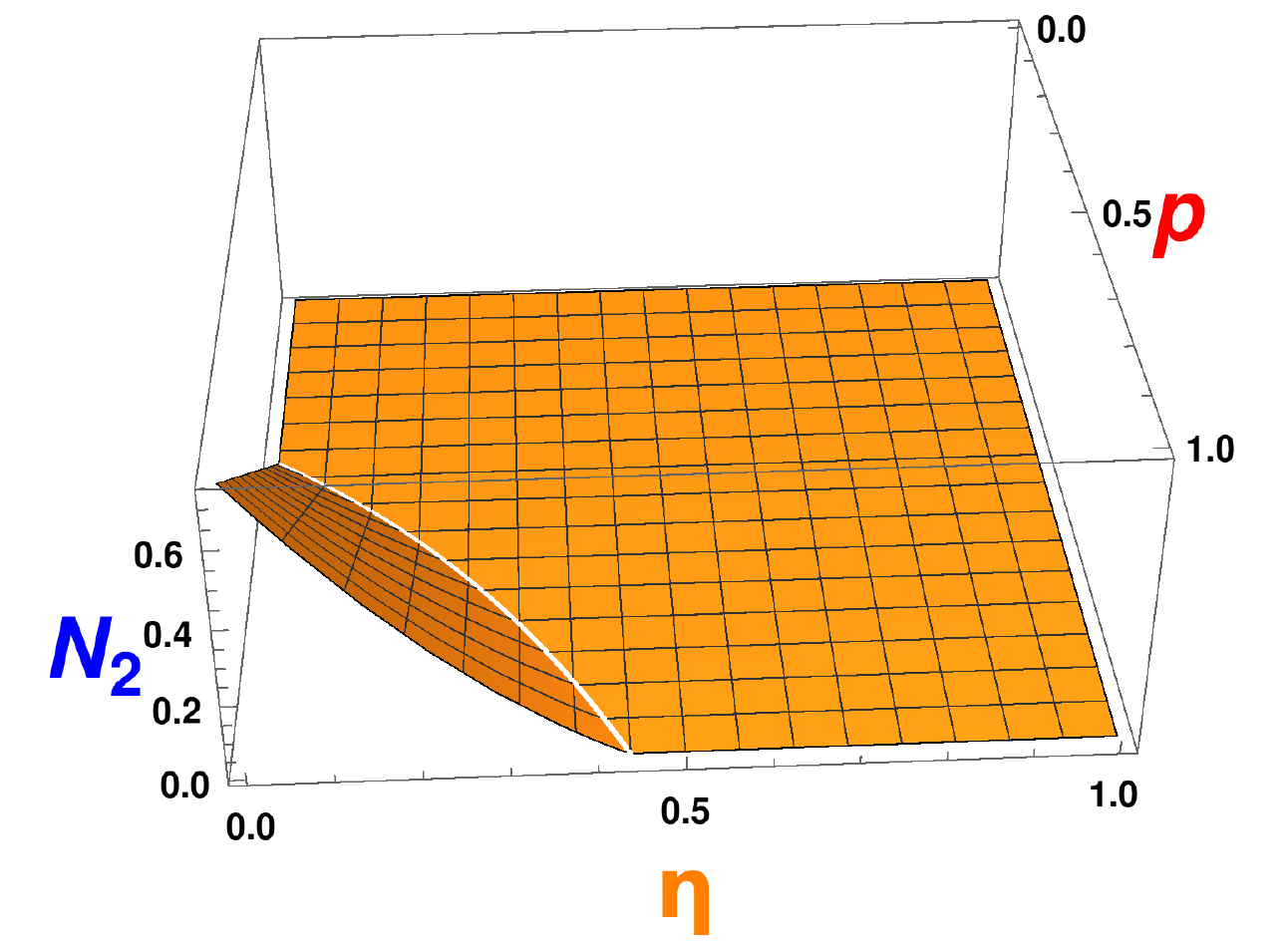}
\caption{(Color online) The figure shows the change of 
the binegativity ($N_2$) of 
initial state $\rho_{EW}$ versus the mixing parameter $p$ and the channel parameter $\eta$ for $\alpha=0.4$ due to  
the action of AD channel on both the particles of the state. 
The effect of the AD channel on both particles is more than 
the single particle one. (Note that the the concurrence ($C$) and the negativity ($N$) behave similarly.)}\label{AD2}
\end{figure}

\begin{figure}[h]
\centering
\includegraphics[scale=0.6]{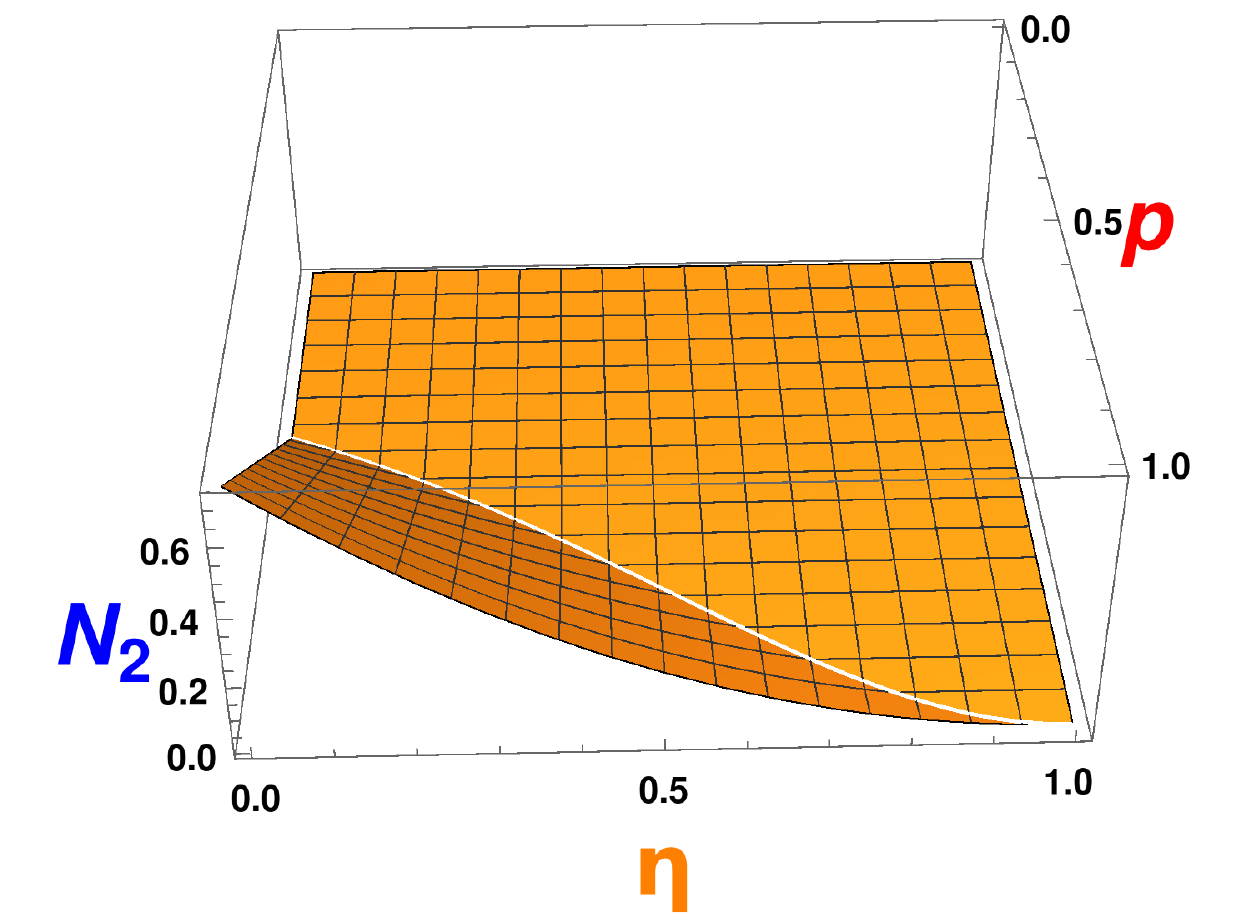}
\caption{(Color online) The plot shows the change of 
the binegativity ($N_2$) of 
initial state $\rho_{EW}$ versus the mixing parameter $p$ and the channel parameter $\eta$ for $\alpha=0.4$ under   
the action of PD channel on both the particles of the state. 
(Note that the the concurrence ($C$) and the negativity ($N$) behave similarly.)}\label{PD2}
\end{figure}

\begin{figure}[h]
\centering
\includegraphics[scale=0.7]{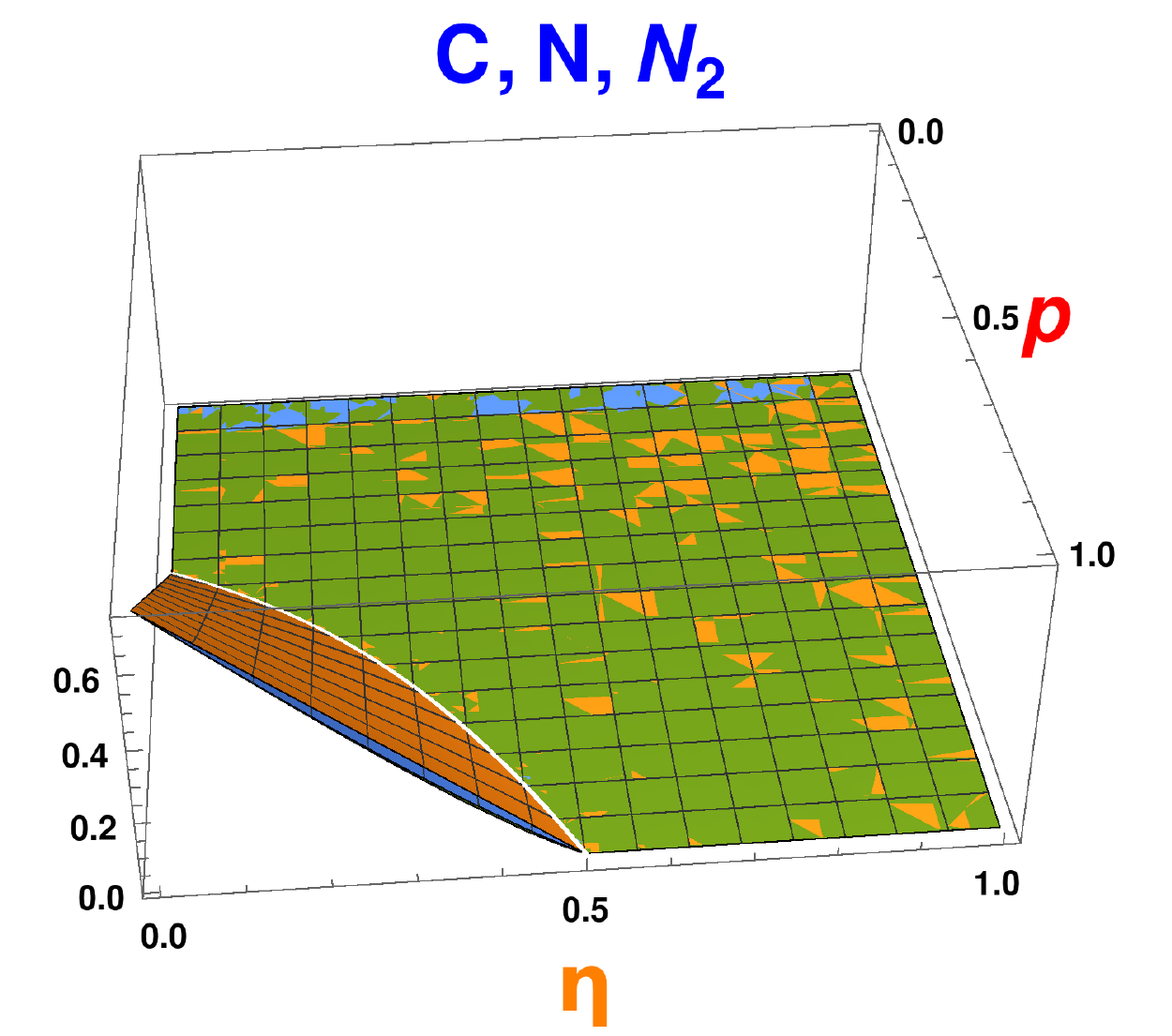}
\caption{(Color online) The graph depicts the change of the concurrence ($C$), 
the negativity ($N$), and the binegativity ($N_2$) of 
initial state $\rho_{EW}$ versus the mixing parameter $p$ and the channel parameter $\eta$ for $\alpha=0.4$ due to  
the action of DP channel on one particle of the state. The negativity (blue curve) and the binegativity 
(green curve) has been affected more due the action of the channel.}
\label{DP1}
\end{figure}

\begin{figure}[h]
\centering
\includegraphics[scale=0.6]{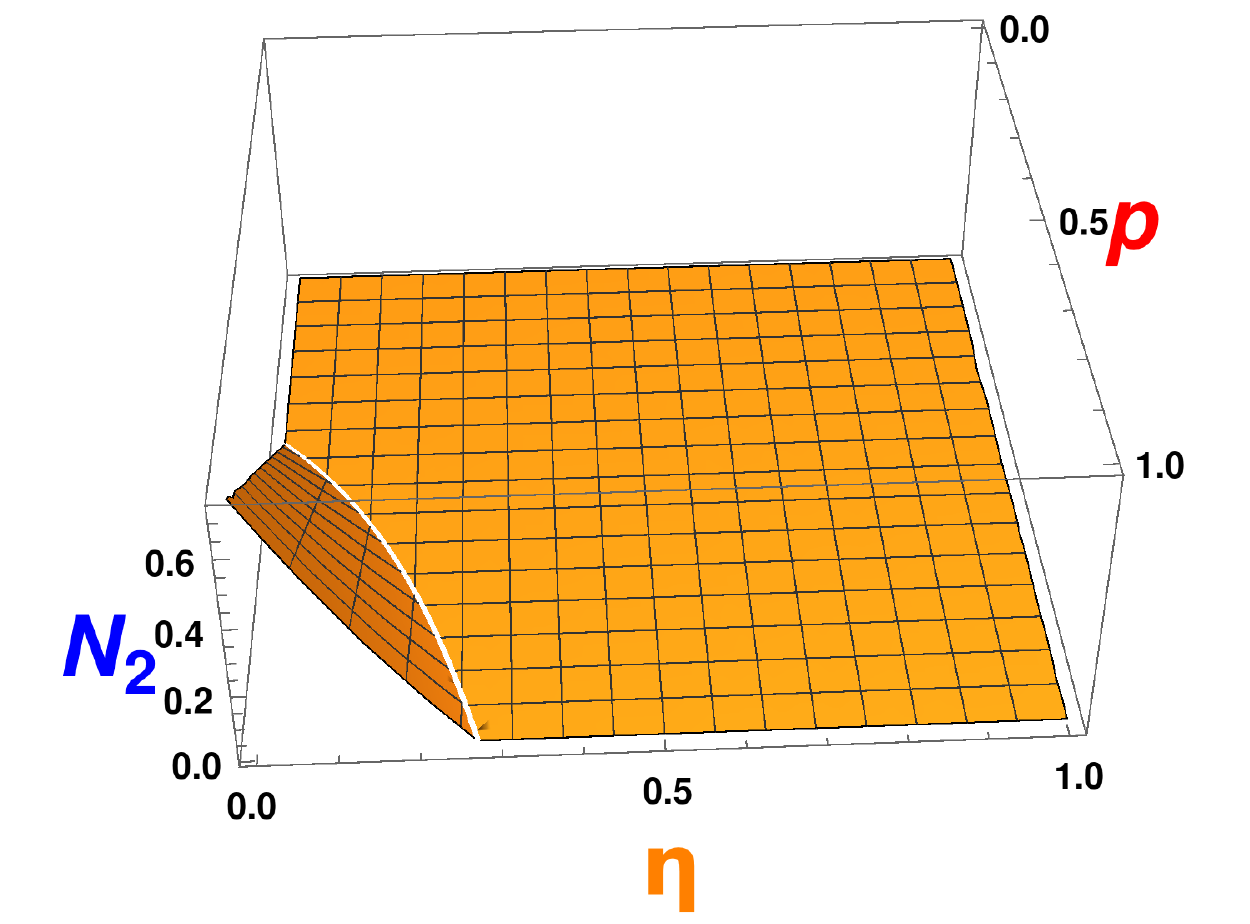}
\caption{(Color online) The graph shows the change of 
the binegativity ($N_2$) of 
initial state $\rho_{EW}$ versus the mixing parameter $p$ and the channel parameter $\eta$ for $\alpha=0.4$ when 
the DP channel is acting on both the particles of the state. 
The effect of the DP channel on both the particles is more than 
the single particle one. 
(Note that the the concurrence ($C$) and the negativity ($N$) behave similarly.)
}\label{DP2}
\end{figure}

The entanglement of the initial state $\rho_{EW}$ captured by the concurrence, the negativity and the binegativity 
are same initially, i.e., direct calculations shows that 
\begin{eqnarray}
 C(\rho_{EW})&=&N(\rho_{EW})= N_2(\rho_{EW})\nonumber\\&=&2\max[0,|p\alpha\beta^*|-\frac{(1-p)}{4}].
\end{eqnarray}
Now under the quantum noisy channels, the entanglement of the initial state will decay with the increase of noise 
parameter $\eta$ as depicted in Figs.(\ref{AD1} -- \ref{DP2}). 

{\em AD channels}.- Due to the application of AD channel on first particle, the state $\rho_{EW}$ evolves to
\begin{equation}
 \rho_{f}^{AD}=\begin{pmatrix}
\ell_++p|\alpha|^2 & 0 & 0 & \epsilon\alpha^*\beta \\
0 & \ell_++p\eta |\beta|^2 & 0 & 0 \\
0 & 0 & \ell_- & 0 \\
\epsilon\alpha\beta^* & 0 & 0 & \ell_-+m|\beta|^2 \\
\end{pmatrix},
\label{finalAD1}
\end{equation}
where $\ell_{\pm}=\frac{(1-p)(1\pm \eta)}{4}$, $\epsilon=p\sqrt{1-\eta}$, and $m=p(1-\eta)$. The direct 
calculation shows that the the concurrence, the negativity and the binegativity are,
\begin{eqnarray}
 C&=&2\max[0,\epsilon|\alpha\beta^*|-\sqrt{\ell_-(\ell_++p\eta|\beta|^2)}],\nonumber\\
 N&=&\max[0, L-(\ell_++\ell_-+p\eta|\beta|^2)],\nonumber\\
 N_2&=&\frac{N}{2}\left(1+2\left|\frac{\epsilon\alpha\beta^*(\vartheta-L)}{4\epsilon|\alpha\beta^*|^2
 +\vartheta(\vartheta-L)}\right|\right),\label{ADexp}
\end{eqnarray}
where $\vartheta=\ell_+-\ell_-+p\eta|\beta|^2$ 
and $L=\sqrt{\vartheta^2+4\epsilon|\alpha\beta^*|^2}$. 

The effect of AD channel on the first particle of the 
state $\rho_{EW}$ is shown in Fig.(\ref{AD1}). It reveals that the concurrence is more robust than the other two while 
binegativity is affected more. If we closely look at the mathematical expressions,
we find that the concurrence, the negativity and the binegativity have different functional behavior with respect 
to the noisy parameter $\eta$ for this case, i.e., 
Eq.(\ref{ADexp}) indicates that $C\sim \sqrt{\eta}$ whereas both
$N$ and $N_2$ have quadratic dependence on $\eta$.

However, when we apply the AD channel on both particles of the state, the above trait vanishes and all the 
entanglement measures behave similarly. In this case the state $\rho_{EW}$ transforms to
\begin{equation}
 \rho_{f}^{AD}=\begin{pmatrix}
s & 0 & 0 & m\alpha^*\beta \\
0 & v & 0 & 0 \\
0 & 0 & v & 0 \\
m\alpha\beta^* & 0 & 0 & (r+p|\beta|^2)(1-\eta)^2 \\
\end{pmatrix},
\label{finalAD2}
\end{equation}
where $s=r(1+\eta)^2+p(|\alpha|^2+\eta^2|\beta|^2)$, $r=\frac{(1-p)}{4}$, and $v=(1-\eta)[\ell_++p\eta^2 |\beta|^2]$. One can directly calculate the concurrence, the negativity and the negativity, i.e., 
\begin{eqnarray}
 C&=&N=N_2\nonumber\\&=&2\max[0, \frac{\epsilon^2}{p}|\alpha\beta^*|-(1-\eta)(\ell_++p\eta^2 |\beta|^2)].
\end{eqnarray}
Hence, all the considered entanglement quantifiers have similar dependence on the noise parameter $\eta$. Furthermore, in this case, the 
decaying of entanglement as captured by the concurrence, the negativity and the binegativity is more as shown in Fig.(\ref{AD2}). 

{\em PD channels}.- The action of PD channel on the single particle as well as on the both particles of the state $\rho_{EW}$ 
will led to the following state,
\begin{equation}
 \rho_{f}^{PD}=\begin{pmatrix}
r+p|\alpha|^2 & 0 & 0 & \frac{m^i}{p}\alpha^*\beta \\
0 & r & 0 & 0 \\
0 & 0 & r & 0 \\
\frac{m^i}{p}\alpha\beta^* & 0 & 0 & r+p|\beta|^2 \\
\end{pmatrix}.
\label{finalPD2}
\end{equation}
The index $i=1$ means PD is applied on single particle and $i=2$ implies PD has been applied 
on both particles.

The effect of PD channel for both the cases are almost similar except for the decay rate (see Figs.(\ref{PD2})). All 
three measures of entanglement decays more rapidly when PD channels are applied to both the particles of the 
state $\rho_{EW}$ because
\begin{eqnarray}
 C&=&N=N_2\nonumber\\&=&2\max[0, \frac{m^{i}}{p^{i-1}}|\alpha\beta^*|-\frac{(1-\eta)}{4}].
\end{eqnarray}
Therefore, analytical results indicate that these entanglement measures have linear dependence on $\eta$ for 
one sided PD channel but has quadratic dependence on $\eta$ for both side PD.

{\em DP channels}.- After the application of DP channel on first particle, the state $\rho_{EW}$ will transform to 
\begin{equation}
 \rho_{f}^{DP}=\begin{pmatrix}
r+pt_2|\alpha|^2 & 0 & 0 & pt_4\alpha^*\beta \\
0 & \Theta_{\beta} & 0 & 0 \\
0 & 0 & \Theta_{\alpha} & 0 \\
pt_4\alpha\beta^* & 0 & 0 & r+pt_2|\beta|^2 \\
\end{pmatrix},
\label{finalDP1}
\end{equation}
where $t_j=1-\frac{j\eta}{3}$ and $\Theta_{x}=r+p\eta |x|^2$ with $j=2,4$. 
Then one can calculate the concurrence, the negativity and the binegativity which are respectively,
\begin{eqnarray}
 C &=& 2 \max [0, |\omega|-\sqrt{\Theta_{\alpha} \Theta_{\beta}}],\nonumber\\
 N &=&\max [0,\Upsilon-(\Theta_{\alpha}+ \Theta_{\beta})],\:\: \mbox{and}\nonumber\\
 N_2 &=&\frac{N}{2}\left(1+\frac{|2\omega(\Theta_{\beta}-\Theta_{\alpha}-\Upsilon))|}
 {|4|\omega|^2+(\Theta_{\beta}-\Theta_{\alpha})(\Theta_{\beta}
 -\Theta_{\alpha}-\Upsilon)|} \right),\nonumber
\end{eqnarray}
where $\omega=pt_4\alpha^*\beta$ and $\Upsilon=\sqrt{(\Theta_{\beta}-\Theta_{\alpha})^2+4|pt_4\alpha^*\beta|^2}$. All these entanglement 
quantifiers behave almost similarly although their analytical expressions are quite different 
(see Figs.(\ref{DP1})). From the figure, it is clear that the affect of DP channel is slightly more on the 
negativity and the binegativity (the blue and green curve respectively).

Whereas the final state will be 
given by the following equation if the DP channels act on both the particles of the state $\rho_{EW}$,
\begin{equation}
 \rho_{f}^{DP}=\begin{pmatrix}
\Delta_{\alpha\beta} & 0 & 0 & \kappa \\
0 & \delta & -\xi\alpha^*\beta  & 0 \\
0 & -\xi\alpha\beta^* & \delta & 0 \\
\kappa^* & 0 & 0 & \Delta_{\beta\alpha} \\
\end{pmatrix},
\label{finalDP2}
\end{equation}
where $\delta=r+\frac{2}{9}p\eta(3-2\eta)$, $\tau=\frac{p}{9}(9-24\eta+14\eta^2)$, $\varsigma=\frac{p}{9}(1-\mi)\eta^2$, 
$\xi=\frac{2}{9}p\eta^2$, $\kappa=\varsigma\alpha\beta^* +\tau\alpha^*\beta$, 
and $\Delta_{xy}=r+pt_2^2|x|^2+2\xi|y|^2$ with $\mi=\sqrt{-1}$. The analytical expression for the considered 
entanglement quantifiers (see the Appendix.\ref{EW-state}) are
\begin{eqnarray}
 C=N=N_2=2\max[0, |\kappa|-\delta].
 \label{finalDP2exp}
\end{eqnarray}
The Eq.(\ref{finalDP2exp}) simplifies because of the fact that 
$|\kappa|> \delta$ holds (We have checked it numerically.).
Hence, all quantifiers have 
similar dependence on the noise parameter see Fig.(\ref{DP2}).

Under DP channel all these measures are finding
the most fragile (see Figs.(\ref{DP1}, \ref{DP2})). This is because under DP channels decoherence 
effect is the most. This can be perceived from the analytic expressions of these bonafied measures under 
one sided DP as well as both sided DP. The entanglement measures considered here behave similarly under both
one sided DP and both sided DP while in the later case decay rate is more. Although we have considered $\alpha=0.4$ for 
numerical depiction, their behavior remain same for any value of $\alpha$.

\section{discussions}
Among the existing entanglement measures, only the concurrence and the negativity are analytically 
computable for arbitrary two qubit states. We have added a new member to this club: the binegativity.
In this paper, we discuss that the binegativity 
might be considered as a faithful measure of entanglement 
for two qubit states. 
This measure coincide with 
the concurrence and the negativity for pure two qubit states. Note that like the negativity, the binegativity 
is not an additive, 
i.e., $N_2(\rho_1\otimes \rho_2)\neq N_2(\rho_1)+N_2(\rho_2)$, where $\rho_i$s are two qubit density matrices. It also 
induces different entanglement orderings among two qubit mixed states \cite{bineg}, which might have an important 
impact on the resource theory of entanglement. 
Therefore, the binegativity is an important 
quantity in resource theoretic perspective \cite{entangr,Mejo}. 

We study the behavior of the binegativity under AD, DP and PD channels 
for the state $\rho_{EW}$ and show that it decreases monotonically with the noise 
parameter $\eta$. We compare the behavior of the binegativity with the concurrence and the negativity 
and find that the binegativity is behaving quite similar to the negativity. Our analysis support the 
conjecture that the binegativity might be a monotone \cite{bineg}. We hope our findings may help in understanding the entanglement structure 
of two qubit mixed states.

\emph{Acknowledgement}:  Natasha acknowledges the local hospitality of HRI, Allahabad.
We thank anonymous referee for 
his insightful comments. For numerics, we took the help of \cite{QET9}.

\appendix
\section{Expressions of $C$, $N$, $N_2$ for the state in Eq.(\ref{finalDP2})}\label{EW-state}

The state in Eq.(\ref{finalDP2}) is of the form 
\begin{equation}
 \rho_{f}=\begin{pmatrix}
a & 0 & 0 & e \\
0 & b & c & 0 \\
0 & c^* & b & 0 \\
e^* & 0 & 0 & d \\
\end{pmatrix},
\end{equation}
where $a,b,d\geq 0$. The concurrence, the negativity and the binegativity of the state are 
\begin{eqnarray}
 C(\rho_{f})=2\max[0,|c|-\sqrt{ad},|e|-b],\nonumber\\
 N(\rho_{f})=\left\{
     \begin{array}{@{}l@{\thinspace}l}
       \theta-(a+d)  &: \text{if}\: a+d<\theta,\\
       2(|e|-b) &: \text{if}\: b< |e|, \\
     \end{array}
   \right.\label{expConc1}
\end{eqnarray}
\begin{eqnarray}
   N_2(\rho_{f})=\left\{
     \begin{array}{@{}l@{\thinspace}l}
       \frac{N(\rho_{f})}{2}\left(1+\left|\frac{2c}{\theta}\right|\right)  &: \text{if}\: a+d<\theta,\\
       N(\rho_{f}) &: \text{if}\: b< |e|, \\
     \end{array}
   \right.\label{expConc2}
\end{eqnarray}
where $\theta=\sqrt{(a-d)^2+4|c|^2}$.

\end{document}